\documentclass[twocolumn,floatfix, nofootinbib,prd,preprintnumbers,showpacs,showkeys,superscriptaddress,longbibliography]{revtex4-1}

\usepackage{ulem}
\usepackage{comment}
\usepackage[mathscr]{euscript}
\usepackage{amsmath}
\usepackage{graphicx}
\usepackage{dcolumn}
\usepackage{bm}
\usepackage{epsfig}
\usepackage{amssymb,latexsym,mathrsfs}
\usepackage{graphicx}
\usepackage{color}
\usepackage{hyperref}
\usepackage{float}
\usepackage{diagbox}

\usepackage{tikz}

\usepackage{amsmath}
\usepackage{csquotes} 
\newcommand*\diff{\mathop{}\!\mathrm{d}}

\hypersetup{
    colorlinks=true,
    linkcolor=red,
    citecolor=blue,
} 

\usepackage{xcolor}

\colorlet{blue}{black}

\usepackage{amsmath}
\usepackage{amssymb}
\usepackage{subfigure}
\usepackage{hyperref}
\usepackage{url}
\usepackage{xcolor}
\usepackage{color}
\definecolor{amaranth}{rgb}{0.9, 0.17, 0.31}
\definecolor{purple(munsell)}{rgb}{0.62, 0.0, 0.77}
\definecolor{americanrose}{rgb}{1.0, 0.01, 0.24}
\definecolor{palatinateblue}{rgb}{0.15, 0.23, 0.89}
\definecolor{royalblue(web)}{rgb}{0.25, 0.41, 0.88}
\definecolor{hanpurple}{rgb}{0.32, 0.09, 0.98}
\definecolor{beaublue}{rgb}{0.74, 0.83, 0.9}
\definecolor{carminered}{rgb}{1.0, 0.0, 0.22}
\definecolor{brightpink}{rgb}{1.0, 0.0, 0.5}
\definecolor{vividviolet}{rgb}{0.62, 0.0, 1.0}
\hypersetup{ linktoc=all,
    colorlinks, linkcolor={palatinateblue},
    citecolor={brightpink}, urlcolor={amaranth}}

\newcommand{\be}{\begin{equation}}
\newcommand{\ee}{\end{equation}}
\newcommand{\bs}{\begin{split}} 
\newcommand{\bea}{\begin{eqnarray}}
\newcommand{\eea}{\end{eqnarray}}

\newcommand{\bes}{\begin{subequations}}
\newcommand{\ees}{\end{subequations}}

\newcommand{\bo}{\raise-1mm\hbox{\Large$\Box$}}

\newcommand{\bd}{\boldsymbol}

\begin{document}

\title{An advanced undergraduate derivation of acceleration thermality}
%

\author{Michael R.R. Good}\,
\email{muon@asu.edu}
\affiliation{Physics Department \& Energetic Cosmos Laboratory, Nazarbayev University,\\
Astana 010000, Qazaqstan.}
\affiliation{Leung Center for Cosmology \& Particle Astrophysics,
National Taiwan University,\\ Taipei 10617, Taiwan.}
\affiliation{Beyond Center for Fundamental Concepts in Science, Arizona State University,\\
Tempe AZ 85287, USA.}

\begin{abstract} 
The thermal radioactivity of beta-decay photons, described by a 1D Planck distribution, can be modeled as classical radiation emitted by an accelerated electron.  Here, we present the basics of the {\color{blue}{out-of-equilibrium}} computation to illustrate acceleration thermality. Suitable for advanced undergraduate calculations, we demonstrate that an exactly soluble non-uniformly accelerated trajectory enables spectral analysis of the emitted {\color{blue}{photons}}, facilitates time evolution, and reveals Planckian {\color{blue}{radiation}}. 

\end{abstract} 

\date{\today} 

\maketitle

\tableofcontents


\section{Introduction}

The relationship between acceleration and temperature is a widely recognized result of relativity and quantum theory \cite{Davies:1974th,unruh76}, often discussed in the context of the equivalence principle and phenomena such as black hole radiation \cite{Hawking:1974sw}. For decades, accelerating systems have provided a fascinating framework for exploring thermality and quantum effects, offering valuable opportunities to deepen our understanding of these areas \cite{Birrell:1982ix,Fabbri,Parker:2009uva}.

Interestingly, in classical and quantum physics, understanding how objects like mirrors\footnote{Idealized moving boundaries in spacetime that reflect quantum fields, often used to model particle creation effects in mathematically simplified settings such as (1+1)-dimensional spacetime.} \cite{moore1970quantum,DeWitt:1975ys,Davies:1976hi,Davies:1977yv,Fulling:1972md}, or electrons \cite{Ievlev:2024zai,Ievlev:2023xzv,Lin:2024ihr,Mujtaba:2024vqy,Good:2022eub,Good:2023ncu,Good:2023hsv,physics5030050,Ievlev:2023akh,Bell:1983}
{\color{blue}{emit radiation with a Planckian spectral distribution while accelerating}} provides a useful perspective on the relationship between acceleration and temperature. Consider classical electrodynamics, where moving point charge radiation is treated in widely-used textbooks; see, e.g., \cite{Morin_Purcell_Electrodynamics,Griffiths:1492149,Jackson:490457,Zangwill:1507229}. 
Here, as it currently stands, thermal radiation from an ordinary, non-uniformly accelerated moving point charge 
has not been treated for undergraduates, leaving students without a good resource to consult for a proper understanding. 

{\color{blue}{This paper presents an advanced undergraduate derivation of a Planckian spectral distribution from a moving point charge.}} Our approach provides a complete yet accessible treatment, leveraging basic classical electromagnetic theory and carefully deriving thermal emission from non-uniform acceleration.  We present a straightforward, step-by-step computation that yields a direct relationship between {\color{blue}{the Planckian radiation, thermal in the spectral sense,}} and an accelerating point charge, offering students better insight into the motion responsible for an exact, analytic, Planck-distributed photon spectrum from a physically motivated trajectory. 

We elucidate the key concepts underlying only the relevant physical quantities of interest by systematically developing the necessary undergraduate-level theoretical computations. Building upon this foundation, we rigorously derive the spectrum and energy of the thermal radiation emitted by the moving charge, providing an example of acceleration-thermality.  {\color{blue}{Here and throughout, ``thermal'' denotes the Planckian spectral form of the emitted radiation, with an effective temperature inferred from that spectrum; it does not imply statistical thermodynamic equilibrium.}}

Our approach incorporates an illustrative integral calculation and highlights the less well-known fact that $\hbar$ appears in all classical acceleration-temperature equations. 
{\color{blue}{The result, consistent with a recent experimental observation in radiative beta decay, supplements students' understanding of the fundamental physics of relativistic motion, Planck's law, and acceleration radiation.}}  Via explicit computation, we hope it helps nurture an intuitive classical background context of acceleration-thermality. 

In Sec.~\ref{sec:elements}, we introduce only the electrodynamic ingredients for computing the radiation spectrum emitted by a moving point charge. We then illustrate, in general, how to calculate the energy using the spectrum. Then, with all the general ingredients prepared, we specialize via a particular non-uniformly accelerated trajectory relevant to beta decay in Sec.~\ref{sec:thermal} to compute the Planck spectral distribution.  In Sec.~\ref{sec:temperature}, we introduce a defining temperature using this Planck spectrum, emphasizing its connection to the acceleration scale and highlighting the one-dimensionality of the energy spectrum. By attending to 1D Planck's law, a reader gains a precise understanding of the temperature of the semi-classical photon spectrum, offering a perspective on the dimensional differences encompassed by thermality. Sec.~\ref{sec:alt} leverages a few approximate schemes, including the non-relativistic regime, infrared frequency, and instantaneous motion. Here, a 3D spectrum is derived. Sec .~\ref {sec:concl} is the conclusion, where we reiterate that the radiation emitted by an accelerating beta particle has a temperature, illustrating a relationship between acceleration and temperature.  

We use SI units in Sec.~\ref{sec:elements}-\ref{sec:temperature}. For the alternative derivations in Sec.~\ref{sec:alt}, we use natural units.

\section{Elements of Electromagnetic Radiation}
In this section, we obtain three main electrodynamic objects of interest: the spectral distribution $\frac{\diff I(\omega)}{\diff \Omega}$, the spectrum $I(\omega)$, and the total energy $E$, given by Eq.~(\ref{specdis}), Eq.~(\ref{IOmega}), and Eq.~(\ref{totenergy}), respectively. 
\label{sec:elements}
\subsection{Moving Point Charge}
In this first subsection, we start with the general expression for the current density of a moving point charge as a function of frequency.  The current density of a point particle with charge $e$, moving along the $z$-axis, is
\be j_z(\bd{r},t) = e \dot{z}(t) \cdot \delta(\bd{r}-z(t)),\ee
where $z(t)$ expresses the straight-line motion of the electron with velocity $\dot{z}(t)$ and observation point $\bd{r}$ in the radial direction pointed away from the origin. The Dirac delta function $\delta(\bd{r}-z(t))$ is the signature defining characteristic of a moving point particle. 

Let us move to frequency space by designating $j_z( \omega,k_z)$ the Fourier transform of the current density,
\begin{equation}
	j_z (\omega, k_z) 
		=  \int\limits_{- \infty}^{\infty} \int\limits_{- \infty}^{\infty}\, j_z(\bd{r},t) \, e^{i \varphi} \diff^3{r} \diff{t},
\label{asdf}
\end{equation}
where the phase is {$\varphi \equiv \omega t - \bd{k} \cdot  \bd{r} $, and with help from the delta function, we obtain,
\begin{equation}
	j_z (\omega, k_z) 
		= e\, \int\limits_{- \infty}^{\infty}  \, \dot{z}(t) \, e^{i \varphi} \diff{t}.
\end{equation}
This simplifies by canceling out the differential $\diff{t}$'s, substituting the phase, $\varphi = \omega t - k_z z(t) $, and re-expressing $z$ as the independent integration variable, {\color{blue}{by using $\dot z(t)\,\diff{t}=\diff{z}$ which converts the time integral into an integral over position along the trajectory,}}
\begin{equation}
	j_z (\omega, k_z) 
		= e\, \int\limits_{0}^{\infty} \, \textrm{exp}[i (\omega t(z) - k_z z)] \diff{z},\label{currentjz}
\end{equation}
where the limits of the integration will start at $z=0$ because the motion that exhibits thermality will be limited to the positive z-axis, and $t(z)$ is the $z$-dependent function of time. In spherical coordinates, $k_z = \omega \cos\theta$. 

In the following subsection, we will use this expression of the current density as a function of frequency, Eq.~(\ref{currentjz}), to find the spectral distribution of radiation emitted by the point charge along $t(z)$.  
\subsection{Spectral Distribution}
Here, we show how to derive the spectral distribution formula using the current density, Eq.~(\ref{currentjz}).  The expression that specifies the amount of energy radiated into a unit solid angle, $\diff{\Omega}= \sin\theta \diff\theta \diff\phi$, in the direction of $\bd{k} = (\omega/c)\hat{r}$ in a unit frequency interval centered on $\omega$ is
\be \frac{\diff I(\omega)}{\diff \Omega} = \frac{\mu_0 \omega^2}{16\pi^3 c} |\bd{\hat{r}} \times j_z(\omega, k_z)|^2,\ee
See, for e.g., Eq. (23.87) of Zangwell \cite{Zangwill:1507229} or Eq. (14.70) of Jackson \cite{Jackson:490457}).
This effectively simplifies to squaring the current density:
\begin{equation}
    \frac{\diff I(\omega)}{\diff \Omega} 
		= \frac{\mu_0 \omega^2}{16 \pi^3 c} \sin^2\theta \, \left|   j_z(\omega, k_z ) \right|^2.
\label{I_Omega_Jackson}
\end{equation}
Substitution of the current density, Eq.~(\ref{currentjz}), gives
\be \frac{\diff I(\omega)}{\diff \Omega} 
		= \frac{\mu_0 e^2\omega^2}{16 \pi^3 c} \sin^2\theta \, \left|\int\limits_{0}^{\infty} \, e^{i\varphi} \diff{z}\right|^2, \label{sdphase}\ee
where it can be seen that the main mathematical task boils down to the integral,
\be \int\limits_{0}^{\infty} \, e^{i\varphi} \diff{z} = \int\limits_{0}^{\infty} \, \textrm{exp}[i\omega t(z) - i \omega z\cos\theta]\diff{z}.\label{phaseintegral}\ee
So, in total, the spectral distribution is written as,
\be \frac{1}{\sin^2\theta}\frac{\diff I(\omega)}{\diff \Omega} 
		= \frac{\mu_0 e^2\omega^2}{16 \pi^3 c} \, \left|\int\limits_{0}^{\infty} \, e^{i\omega(t(z) - z\cos\theta)}\diff{z} \right|^2. \label{specdis}\ee
We will introduce the exact trajectory, $t(z)$, as a function of the independent variable $z$ in Section \ref{particulartraj}, Eq.~(\ref{tofz}).  To keep things general, the next subsection demonstrates how to use the spectral distribution to find the spectrum $I(\omega)$ and the total energy $E$. 
 
\subsection{Spectrum and Energy}

The spectrum $I(\omega)$ is found by integrating the spectral distribution over the solid angle. The solid angle, $\diff{\Omega}= \sin\theta \diff\theta \diff\phi$, helps quantify how much a particular surface region captures outgoing light in three-dimensional space. {\color{blue}{The energy spectrum is:}}
\begin{equation}
I(\omega)=\int \frac{\mathrm{d} I(\omega)}{\mathrm{d} \Omega} \diff{\Omega}\label{IOmega} = \int_0^{2\pi}\int_0^{\pi} \frac{\mathrm{d} I(\omega)}{\mathrm{d} \Omega} \sin\theta \diff{\theta}\diff{\phi}.\end{equation}
The energy is found by integrating over the frequency $\omega$, 
\begin{equation}
E = \int_{0}^{\infty} I(\omega) \, \diff{\omega},\label{totenergy}
\end{equation}
allowing us to find the total energy emitted by the spectrum $I(\omega)$ over the entire frequency range.

Overall, these main electrodynamic objects of interest: the spectral distribution, the spectrum, and the total energy, given by Eq.~(\ref{specdis}), Eq.~(\ref{IOmega}), and Eq.~(\ref{totenergy}), respectively, are the most significant electrodynamic ingredients necessary to derive thermal radiation from an accelerating point charge. In the next section, we introduce and motivate a specific, non-uniformly accelerated trajectory and solve for these specific quantities of interest.

\section{Trajectory-Specific Electromagnetic Radiation} 
\label{sec:thermal}
In this section, we compute thermal radiation from an accelerated electron.  We now have all the necessary ingredients to use a particular equation of motion to describe the electron's trajectory. 
\subsection{The Particular Trajectory}\label{particulartraj}
Uniform acceleration is an intuitive and simple trajectory that relates a constant acceleration to a steady temperature, as in the Unruh effect \cite{unruh76}; see also e.g., various uniformly accelerated trajectory temperatures \cite{Good:2020hav}.  {\color{blue}{Here, however, the thermality needs to be purely kinematic rather than statistical: the present problem is classical radiation emitted by an out-of-equilibrium accelerated charge, not the equilibrium response of a detector in the Minkowski vacuum.}}  Thus, uniform acceleration is unsuitable for an accelerating electron due to two key issues:
\begin{enumerate}
    \item Maintaining uniform acceleration indefinitely would cause the electron's speed to asymptotically approach the speed of light, requiring infinite energy to sustain.
    \item {\color{blue}{The uniformly accelerated}} electron's asymptotic approach to a horizon, defined by the last light ray to reach it, leads to an infinite amount of radiation energy observed at a distance.

\end{enumerate}
The second can be addressed by choosing a different worldline trajectory that ensures the asymptotic approach to the speed of light is timelike (thus preventing the formation of a horizon).  A natural choice is an exponential function of radial distance in terms of retarded time \cite{Good:2022eub}, which is a good illustration of a time-like (not light-like) horizonless approach to the speed of light:
\be r = \frac{c^2}{\kappa} e^{\frac{\kappa}{c} u },\qquad u = t - \frac{r}{c}.\label{exptraj} \ee
However, despite being horizonless, this trajectory still suffers from the first issue (infinite energy is required to push a massive particle to the speed of light). Nevertheless, by introducing $s < c$, a final speed parameter, we can effectively `slow down' the electron and prevent an asymptotic approach to the speed of light.  The modified trajectory becomes\footnote{{\color{blue}{This trajectory is not arbitrary. Within the M\"obius symmetry group, it is a particularly natural choice satisfying the physical conditions appropriate for an accelerating electron: subluminal velocity at all times and asymptotically vanishing acceleration. Other M\"obius-transformed trajectories need not share these features, making this example especially convenient for the present analysis.}}} 
\be r = \frac{c s}{\kappa} e^{\frac{\kappa}{c} u_s },\qquad u_s = t - \frac{r}{ s}.\label{exptraj} \ee
The final speed of the electron is $0<s<c$, while $\kappa > 0$ is a constant with units of acceleration. The $u_s = t - r/s$ is a stretched space retarded time parameter. {\color{blue}{For the present construction, this choice is physically motivated by its subluminal, horizonless asymptotics and pedagogically useful because it leads to a closed-form expression for $t(z)$, making the origin of the exact spectrum transparent.}}

With this alternative, non-uniformly accelerated trajectory, a constant (analogous to uniform proper acceleration) of the equation of motion Eq.~(\ref{exptraj}) is 
\be \mathcal{C} \equiv \frac{r''(u_s)}{r'(u_s)}c = \kappa,\label{peel}\ee
where the prime indicates a derivative with respect to the argument, $u_s$ {\color{blue}{(see a similar object known as the peel acceleration, e.g., \cite{Bianchi:2026xoi,Barcelo:2010pj,Barcelo:2010xk})}}. If we allow the straight-line motion along the z-axis, then $r\to z$.  It is easy to solve Eq.~(\ref{exptraj}) for time $t(z)$ as a function of spatial position $z$,
\begin{equation}
t(z) = \frac{c}{\kappa} \ln \left(\frac{\kappa z}{s c}\right) + \frac{z}{s}.\label{tofz}
\end{equation}
This expression for time, $t(z)$, will be used to solve for the spectral distribution. This is what we do in the following subsection. 
\subsection{The Spectral Distribution} 
We can plug $t(z)$ from Eq.~(\ref{tofz}) into the phase $e^{i\varphi}$ of Eq.~(\ref{sdphase}) to get the spectral distribution.  The integrand in Eq.~(\ref{phaseintegral}) is
\be e^{i\varphi} = \textrm{exp}[i\omega t(z) - \frac{i\omega}{c} z\cos\theta],\ee
and upon substitution of Eq.~(\ref{tofz}) becomes
\be e^{i\varphi} = \left(\frac{\kappa z}{c s}\right)^{\frac{i \omega c}{\kappa}}\textrm{exp}\left[\frac{i\omega z}{s}\left(1- \frac{s}{c}\cos\theta\right)\right] .\ee
{\color{blue}{The key point is that the logarithmic term in $t(z)$ converts the phase into a power law, $(\kappa z/cs)^{i\omega c/\kappa}$, so that the resulting Fourier integral takes the same structural form that produces a Planck factor, common in related moving mirror \cite{Good:2022wpw,carlitz1987reflections,CW2lifetime,Good:2019tnf,walker1985particle,GoodMPLA} and Unruh-type computations, although here, the present setting is classical radiation from an accelerated charge.}}

Integrating  over $\diff{z}$, Eq.~(\ref{phaseintegral}), and complex conjugating, as in Eq.~(\ref{sdphase}), with the help of the identity {\color{blue}{(see e.g., Appendix C, Eq.~(C7), of Ref.~\cite{Ievlev:2023inj}),}}
\begin{equation}
\left|\int_0^{\infty} x^{i \alpha} e^{i \gamma x} \mathrm{~d} x\right|^2=\frac{1}{\gamma^2} \frac{2 \pi|\alpha|}{e^{2|\alpha| \pi}-1},\label{integralidentity}
\end{equation}
gives the exact spectral distribution,
\begin{equation}
    \frac{\mathrm{d} I(\omega)}{\mathrm{d} \Omega}=\frac{\mu_0 c e^2 s^2 \sin ^2 \theta}{16 \pi^3(c-s \cos \theta)^2} \frac{2 \pi c \omega / \kappa}{e^{2 \pi c \omega / \kappa}-1}.\label{sdthermal}\end{equation}
Notice the Planck factor, $\mathcal{N}_\omega \equiv (e^{2\pi c \omega/\kappa}-1)^{-1}$, in the spectral distribution. {\color{blue}{Let us now find the spectrum, $I(\omega)$, and total energy emitted, $E$, in the next subsection.}}
\subsection{The Spectrum and Energy}
Despite already having the Planck factor in Eq.~(\ref{sdthermal}), completeness compels us to go ahead and integrate over the solid angle $\diff{\Omega}$ using Eq.~(\ref{IOmega}) to obtain the exact spectrum $I(\omega)$.  The result is
\begin{equation}
    I(\omega)=\frac{\mu_0 c e^2}{2 \pi^2}\left(\frac{ \tanh^{-1} s / c}{s/c}-1\right) \frac{2 \pi c \omega / \kappa}{e^{2 \pi c \omega / \kappa}-1} .\label{IwANSWER} \end{equation}
The total energy can be found by integrating over $\diff{\omega}$ using Eq.~(\ref{totenergy}) to obtain
\be E = \frac{\mu_0 e^2 \kappa}{24\pi}\left(\frac{\tanh^{-1}{s/c}}{s/c} -1\right).\label{energytotspecific}\ee
This energy can be compared to the emission energy released during infrared scattering  \cite{Jackson:490457}, see also the energy released to lowest order from braking radiation during beta decay  
\cite{PhysRev.76.365,Zangwill:1507229}.

\section{Trajectory-Specific Photon Temperature}
\label{sec:temperature}
\subsection{Planck Spectrum}

How will we define temperature? We will use the Planck spectrum and its essential connection to temperature. Black holes \cite{Hawking:1974sw}, electrons \cite{Nikishov:1995qs}, 
and resistors \cite{nyquist} emit photons with a Planck distribution factor: 
\begin{equation}
\mathcal{N}_\omega = \frac{1}{e^{\frac{\hbar\omega}{k_B T}}-1}.\label{planckfactor}
\end{equation}
Electromagnetic waves with frequency distribution according to this factor, $\mathcal{N}_\omega$, are considered to have a defined temperature, $T$. The same holds true for photons at a given frequency $\omega$. {\color{blue}{Accordingly, the temperature introduced should be understood as a spectral temperature scale associated with Planck’s law, rather than as evidence of a density-matrix, KMS, or detailed-balance description.}}
\begin{figure}[h] 
    \centering 
    \includegraphics[width=0.5\textwidth]{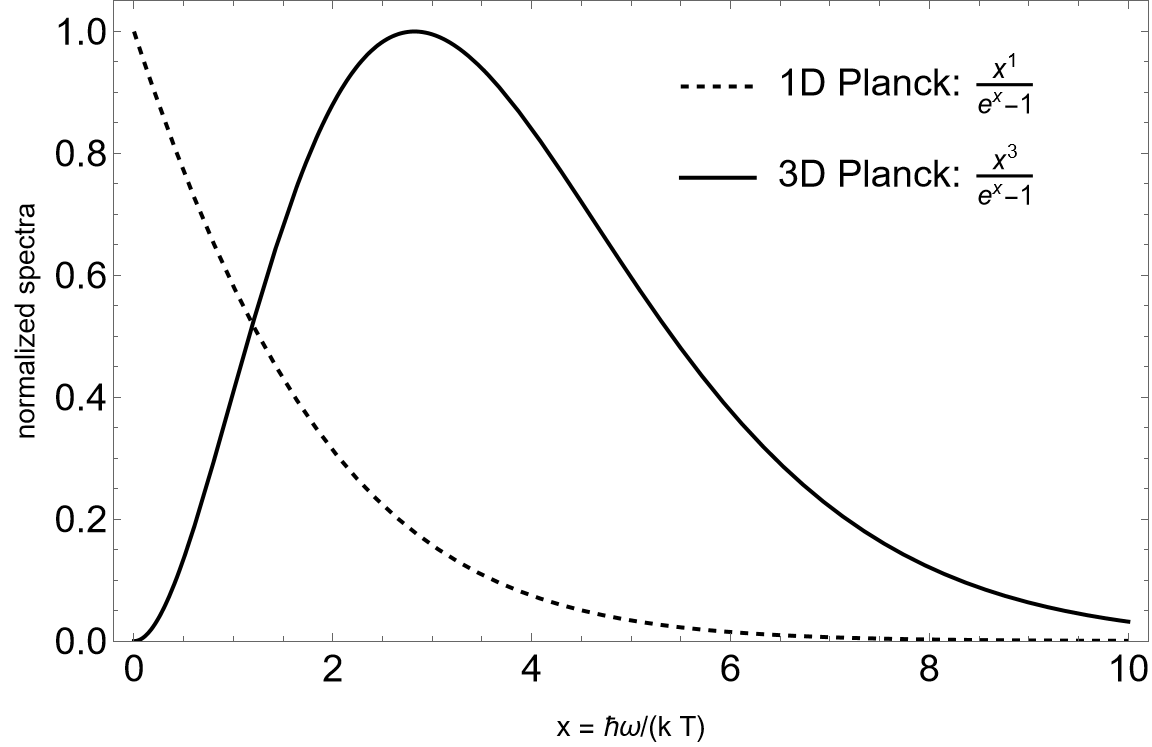}
    \caption{{\color{blue}{Comparison of normalized one-dimensional and three-dimensional Planck spectral shapes, proportional to $x/(e^x-1)$ and $x^3/(e^x-1)$, respectively, where $x=\hbar\omega/k_B T$. The one-dimensional case is infrared-finite and attains its maximum at $\omega=0$, whereas the three-dimensional case peaks according to Wien's displacement law.}}} 
    \label{fig1D}
\end{figure}
The Planck distribution equation precisely defines the meaning of the temperature for the photon spectrum. Notice that the 1D Planck distribution, 
\be I(\omega) \propto \omega \mathcal{N}_\omega, \ee
is distinct from the 3D version,
\be I(\omega) \propto \omega^3 \mathcal{N}_\omega. \ee
{\color{blue}{For visual comparison, Figure~\ref{fig1D} shows the difference between the one-dimensional and three-dimensional Planck spectral shapes.}} The 1D version characterizes the thermal radiation from the accelerated electron (as well as black holes \cite{Bekenstein:2001zoa, Bekenstein:2001tj} and resistors \cite{reif}). Hence, the temperature, which describes the photons emitted by an accelerating electron, is more closely related to photons from a black hole {\color{blue}{(in the 1D sense \cite{Bekenstein:2001tj})}} or the white noise of a 1D resistor \cite{reif} than the photons of the 3D distribution of the Cosmic Microwave Background or an ordinary oven.  This is a dimensional characterization of the energy spectrum, a noteworthy feature of the Planck distribution.

As is known, in 3D systems, the quadratic density of states reflects the isotropic availability of modes in three spatial dimensions, leading to more high-frequency modes than in
1D systems. In 1D systems, the linear scaling of states reflects a limited number of frequency modes, leading to a faster drop in energy density at higher frequencies. However, for specific systems, even in 3D space, the spectral density may scale linearly with frequency and thus, loosely speaking, resemble 1D behavior (this case is no exception). Constraints on the system's effective dimensionality or mode structure, for example, arise from boundary effects, waveguiding, or the creation of symmetries, as in photonic crystals, quantum wires, or surface waves. Thus, asking how the modes are constrained in this accelerating electron case is an interesting question.  As we have seen, the resulting 1D Planck spectrum results from the highly restrictive assumption of rectilinear acceleration (1D) in 3D space, which is responsible for the 1D effective dimensionality.

\
\subsection{Particle Statistics}
In this subsection, we will deviate from the classical approach in previous sections by assuming discrete photons comprise electromagnetic radiation.  The number of particles at a given frequency is given by dividing the spectrum by the quantum of energy for a photon, $\hbar \omega$.
       \be n(\omega) = \frac{I(\omega)}{\hbar \omega}. \label{semiclassical}\ee
{\color{blue}{This step is a semiclassical reinterpretation of the classical energy spectrum, since dividing by $\hbar\omega$ assigns the energy in each mode to discrete photon quanta without invoking a full operator quantization of the radiation field.}} Using $I(\omega)$ in Eq.~(\ref{IwANSWER}) for our specific trajectory, it is clear that the number of photons is proportional to the Planck factor $\mathcal{N}_\omega$:
       \be n(\omega) = \frac{\mu_0 c^2 e^2}{\pi\hbar \kappa }\left(\frac{\tanh^{-1} s/c}{s/c}-1\right) \frac{1}{e^{2\pi c \omega/\kappa}-1}. \label{photons_spectrum}\ee
Recall that $\kappa$ or $s$ are the only parameters specific to the point charge's motion, introduced in Eq.~(\ref{exptraj}).  That is, we keep in mind that $\kappa$ is a constant of the motion with units of acceleration, and $s$ is the final speed of the electron. 

Eq.~(\ref{photons_spectrum}) is a non-classical (i.e., semi-classical) expression of the particle spectrum. {\color{blue}{In particular, as $\omega\to 0$, Eq.~(26) behaves as $n(\omega)\sim \omega^{-1}$, so the integrated particle count has the familiar infrared divergence of a one-dimensional Planck spectrum; however, the corresponding energy spectrum $I(\omega)=\hbar\omega\,n(\omega)$ remains finite in this limit.}} 

The particles in Eq.~(\ref{photons_spectrum}) are the photons emitted from the accelerated electron.  The $\hbar$ in the denominator appears because we introduced a quantum of energy.  Conversely, in the following subsection, we will see how $\hbar$ also appears in the temperature because of the introduction of the Kelvin scale. 
\subsection{Temperature Scale}

A temperature scale can be expressed in any convenient unit system ideal for the system of interest. The Kelvin scale is the SI unit of temperature.  It used to be defined in terms of the triple point of water; however, now the Kelvin scale (without directly mentioning Joules) is defined in terms of Boltzmann's constant, caesium-133 hyperfine transition frequency, and Planck's constant, \cite{SI}
\be 1\; \textrm{K} = \frac{2\pi \times 1.380649 \times 10^{-23}}{
(6.62607015 \times 10^{-34}) (9192631770)}\frac{\Delta \nu_{\textrm{Cs}} \hbar}{k_B}.\ee
This ultimately, although indirectly, associates Planck's constant with temperature metrology.  Since Planck's constant is often well-associated with quantum theory, and our computation for $I(\omega)$, Eq.~(\ref{IwANSWER}), is classical, we highlight that simply introducing the popular Kelvin scale requires $\hbar$.

Typically, the appearance of $\hbar$ signals a computation that invokes quantum mechanics, but our result for the energy spectral distribution, energy spectrum, and energy [Eq.~(\ref{sdthermal}), Eq.~(\ref{IwANSWER}), and Eq.~(\ref{energytotspecific}) respectively] does not.  Of course, the presence of $\hbar$ in the radiation temperature assumed from the spectral energy quantities does not inherently require a quantum mechanical computation.  The classical approach, associating a temperature with the Planck distribution in the energy spectrum, necessarily includes $\hbar$ only via the use of SI units.  After all, a choice of unit system isn't physical. {\color{blue}{Nevertheless, we keep in mind that the assignment of a photon-number spectrum $n(\omega)=I(\omega)/\hbar\omega$ requires $\hbar$ due to the semiclassical approach.}}

We may use the semiclassical expression Eq.~(\ref{semiclassical}), $n(\omega) = I(\omega)/\hbar\omega$, not just $I(\omega)$ in Eq.~(\ref{IwANSWER}), to associate a temperature to the particles $n(\omega)$ of Eq.~(\ref{photons_spectrum}),
\be T = \frac{\hbar \kappa}{2\pi k_B c}. \label{temp}\ee
This expression is in Kelvin. {\color{blue}{Eq.~(\ref{temp}) has the same functional form as the Unruh temperature, but here it arises from the Planck factor, Eq.~(\ref{planckfactor}), $\mathcal{N}_\omega = (e^{\frac{\hbar\omega}{k_B T}}-1)^{-1}$, generated by the Fourier structure of the non-uniformly accelerated trajectory rather than from uniformly accelerated detector response.}} Eq.~(\ref{temp}) is the temperature of photons emitted by an electron traveling along the trajectory, Eq.~(\ref{exptraj}).

{\color{blue}{The scope of the present treatment is therefore classical kinematics at the level of the radiated Planckian spectrum: it's important to distinguish that this is not a quantum-statistical maximized entropy state, detector response, or thermodynamic equilibrium of the radiation field.}} This concludes the derivation of {\color{blue}{Planckian}} radiation from an accelerated point charge. 
\section{Alternative Derivations}\label{sec:alt}
In this section, we present a few alternative derivations to orient the reader to the context and relevance of the preceding results. For simplicity, we use natural units, $c=\mu_0=\epsilon_0 = k_\textrm{B} = \hbar =1$.

The first subsection derives the thermal spectrum straightforwardly, but by brute force, in the non-relativistic regime.  The second subsection bypasses a frequency-dependent spectrum and presents infrared results, thereby connecting the total energy emitted to the known experimental results for radiative beta decay. The third subsection derives a three-dimensional Planck spectrum (using a trajectory that comes to a stop) that accounts for finite-particle emission.    
\subsection{Non-relativistic Regime}
Here, we present the relevant formulae for photon radiation within the non-relativistic regime. The total emitted photon energy radiated by the accelerated electron during beta decay with acceleration parameter $\kappa$, and asymptotic final speed $s$, is given by, e.g.\  \cite{Ievlev:2023inj},  
\be
E = \frac{e^2 \kappa}{24\pi} \left( \frac{\tanh^{-1}s}{s} - 1 \right) \approx \frac{e^2 \kappa}{24\pi} \left(\frac{s^2}{3}\right),\label{beta_decay_energy}
\ee
which agrees with known experimental results, see e.g., \cite{PhysRev.76.365,Lynch:2022rqx}.

Consider a classical point particle electron moving along the trajectory $z(t)$, \cite{Good:2022eub},
\be
z(t) = \frac{s}{\kappa} \, W\left( e^{\kappa t} \right),\label{betadecay_traj}
\ee
where \( W \) denotes the product logarithm, \( s \) is the final speed of the electron with \( s \ll 1 \), and \( \kappa \) is the acceleration scale. From this classical trajectory, it is straightforward to verify that the total radiated energy computed via the Larmor formula for power, $P = e^2 \ddot{z}^2/(6\pi)$, yields:
\be
E = \frac{e^2}{6\pi} \int_{-\infty}^{+\infty} \ddot{z}^2(t) \, \diff{t} = \frac{e^2 \kappa}{72\pi} s^2.
\ee
Using semi-classical physics, recall the non-relativistic particle spectrum (e.g., the classical analog of Eq.~(59) of \cite{Mujtaba:2024vmf}), each quanta has energy $\omega$:
\be
N(\omega) = \frac{I(\omega)}{\omega}, \quad \text{where} \quad I(\omega) = \frac{e^2}{3\pi} \left| \ddot z(\omega) \right|^2. \label{Iw}
\ee 
Using the mathematical identity, Eq.~(32) of \cite{Mujtaba:2024vmf}, 
\be
\left| \mathcal{F}_\omega \left[ \partial_t^2 \frac{1}{\kappa} W(e^{\kappa t}) \right] \right|^2 = \frac{\omega/\kappa}{e^{2\pi \omega/\kappa} - 1},\ 
\label{1D_law}\ee
where the Fourier transform is defined as
\begin{equation}
    z(\omega) = \mathcal{F}_\omega z(t)= \frac{1}{\sqrt{2\pi}}\int_{-\infty}^{+\infty} \diff{t} \, z(t) e^{-i\omega t}  \,,
\label{fourier_def_2}
\end{equation}
and plugging into Eq.~(\ref{Iw}), we find
\be
I(\omega) = \frac{e^2 s^2}{3\pi} \frac{\omega/\kappa}{e^{2\pi \omega/\kappa} - 1}\ . \label{NR_thermal}
\ee
Consequently, the total particle-energy is found to be
\be
E = \int_{0}^{\infty} I(\omega) \, \diff \omega = \frac{e^2 \kappa}{72\pi} s^2,
\ee
in agreement with the known result, Eq.~(\ref{beta_decay_energy}). 

\subsection{Radiative Beta Decay}

To motivate the {\color{blue}{limited}} relevance of the one-dimensional thermal spectrum, Eqs. ~(\ref{IwANSWER}) and (\ref{NR_thermal}), we note that its form is that of gamma radiation emitted during beta decay, {\color{blue}{and is supported by a recent experimental confirmation}}, see e.g., \cite{Lynch:2022rqx}.   Consider that the thermal angular spectral distribution Eq.~(\ref{sdthermal}), may be approximated at leading order $\omega \to 0$,
\be
\begin{array}{rl}
\displaystyle 
\frac{\mathrm{d}I(\omega)}{\mathrm{d}\Omega}
&= \displaystyle 
\frac{e^2}{16\pi^3} \left(\frac{s \sin\theta}{1-s \cos\theta}\right)^2
\,\frac{2\pi\omega/\kappa}{e^{2\pi\omega/\kappa}-1},
\\[1.2em]
&\simeq \displaystyle 
\frac{e^2}{16\pi^3} \left(\frac{s \sin\theta}{1-s \cos\theta}\right)^2. \label{sd_approx}
\end{array}
\ee
Integrating over the solid angle gives a frequency-independent spectrum,
\be I(\omega) = \frac{ e^2}{2 \pi^2}\left(\frac{ \tanh^{-1} s }{s}-1\right), \ee
whose resulting radiated photon energy is
\be
\begin{array}{rl}
E &= \displaystyle \int_{\omega_{\textrm{min}}}^{\omega_{\textrm{max}}} I(\omega) \diff{\omega},
\\[1.2em] 
&= \displaystyle\frac{ e^2}{2 \pi^2}\left(\frac{ \tanh^{-1} s }{s}-1\right) (\omega_{\textrm{max}}-\omega_{\textrm{min}}). \label{photon_energy_cuts}
\end{array}\ee
which scales precisely as Eq.~(\ref{energytotspecific}), when the $\kappa$ is set to 
\be \kappa = \frac{12}{\pi}(\omega_{\textrm{max}}-\omega_{\textrm{min}}). \label{kappa_scale}\ee
For beta decay, these frequency cut-offs are dictated by the energy available to the photon, namely
\be m_{\textrm{neutron}} - m_{\textrm{proton}}-m_{\textrm{electron}} = \omega_{\textrm{max}}-\omega_{\textrm{min}}.
\ee
Since Eq.~(\ref{temp}), $T = \kappa/2\pi$, one has
\be \Delta M = \frac{\pi^2}{6} T_{\textrm{photon}}. \ee
A couple of alternative approximations may be used to illustrate and derive this angular spectral distribution Eq.~(\ref{sd_approx}). In the following two subsections, we demonstrate the approximate form of Eq.~(\ref{sd_approx}) and thus the total energy emitted by photons during beta decay, Eq.~(\ref{photon_energy_cuts}) [Eq.~(\ref{energytotspecific}) with Eq.~(\ref{kappa_scale})], by specializing via two limiting regimes: zero frequency and zero time.

\subsubsection{Derivation 1: Zero-Freq Limit}
One approximation that computes the leading-order frequency-independent angular distribution of Eq.~(\ref{sd_approx}), where $d^2E/d\omega d\Omega \equiv dI(\omega)/d\Omega$, assumes infrared light only; see the approach used in Jackson \cite{Jackson:490457}. We start this derivation with the general angular distribution formula (see, e.g., Zangwill \cite{Zangwill:1507229}): 
\begin{equation}\frac{\diff I(\omega)}{\diff \Omega} = \frac{e^2}{16\pi^3} \left|\int\displaylimits_{-\infty}^{\infty}dt_r\frac{[\hat{\bd{n}} \times ((\hat{\bd{n}}-\bd{\beta})\times \dot{\bd{\beta}}]}{(1-\beta\cos\theta)^2} e^{i\phi}\right|^2,\label{Iw}\ee 
and use the differential identity:
\begin{equation}\frac{\hat{\bd{n}} \times ((\hat{\bd{n}}-\bd{\beta})\times \dot{\bd{\beta}}}{(1-\beta\cos\theta)^2} = \frac{d}{dt_r} \left[ \frac{\hat{\bd{n}}\times(\hat{\bd{n}}\times \bd{\beta})}{1-\beta\cos\theta} \right],\label{totald}\ee
where the derivatives are evaluated at retarded time.  The distribution is then expressed as
\be \frac{\diff I(\omega)}{\diff \Omega} = \frac{e^2}{16\pi^3} \left|\int\displaylimits_{-\infty}^{\infty}dt_r  \frac{d}{dt_r} \left[\frac{\hat{\bd{n}}\times(\hat{\bd{n}}\times \bd{\beta})}{1-\beta\cos\theta} \right] e^{i\phi}\right|^2.\ee
For low frequencies, $e^\phi \to 1$, where $\phi =\omega t_r -\bd{k}\cdot\bd{r}_0(t_r) \to 0$ for $\omega \to 0$. The integral is now a perfect differential, and one has (assuming the charge is initially at rest):
\be \frac{\diff I(\omega)}{\diff \Omega} = \frac{e^2}{16\pi^3} \left|\frac{\hat{\bd{n}}\times(\hat{\bd{n}}\times \bd{\beta})}{1-\beta\cos\theta}\right|^2.\ee
The numerator simplifies because $|\hat{\bd{r}}\times(\hat{\bd{r}}\times \bd{\beta})|^2 = |\hat{\bd{r}}\times \bd{\beta}|^2$ for any vector $\bd{\beta}$.  At late times, when $\bd{\beta}(t) = \bd{\beta}_\textrm{f}$ and the final magnitude is speed, $s =|\bd{\beta}_\textrm{f}|$, the energy distribution per frequency per solid angle is exactly Eq.~(\ref{sd_approx}):
\begin{equation}\frac{\diff I(\omega)}{\diff \Omega} = \frac{e^2}{16\pi^3} \left(\frac{s \sin\theta}{1-s \cos\theta}\right)^2.\ee

\subsubsection{Derivation 2: Zero-Time Limit}

The computation of the angular distribution in Eq.~(\ref{sd_approx}) can also be done without direct appeal to the low-frequency limit.  It involves strict reliance on a step function trajectory, where the electron is assumed to be initially at rest at $t=0$ and imagined to be violently accelerated to a final constant speed, $s= |\vec{\beta}_{\textrm{f}}|$  
where $0<s<1$,
\begin{equation}
  v(t) =
    \begin{cases}
      s, &  t>0.\\
      0, &  t<0.
     \end{cases}   \label{stepspeed}   
\end{equation}
{\color{blue}{That is, this step-function trajectory is a physically singular idealization: with infinite acceleration in zero time, it is used here only as a limiting mathematical model that makes the problem tractable and exposes the spectral structure of Eq.~(\ref{sd_approx}).}} 

To compute the angular distribution, $d^2E/d\omega d\Omega \equiv dI(\omega)/d\Omega$, without the low-frequency approximation, we instead perform an integration by parts inside the absolute signs of Eq.~(\ref{Iw}) after using Eq.~(\ref{totald}), which gives
\be 
\left.\frac{\hat{\bd{n}}\times(\hat{\bd{n}}\times \bd{\beta})}{1-\beta\cos\theta} e^{i\phi}\right|_{0}^{+\infty}- i\omega \int\displaylimits_{0}^{\infty} d t_r\, \hat{\bd{n}}\times(\hat{\bd{n}}\times \bd{\beta})e^{i\phi},\label{parts}
\ee
where the boundary terms vanish.  Here $\phi =\omega t_r -\bd{k}\cdot\bd{r}_0(t_r)$, where $k=\omega$. Using $\hat{\bd{n}} \approx \hat{\bd{r}}$ as a constant vector means it comes outside the integral. Since, as before, $|\hat{\bd{r}}\times(\hat{\bd{r}}\times \bd{\beta})|^2 = |\hat{\bd{r}}\times \bd{\beta}|^2$ for any vector $\bd{\beta}$, we obtain the angular spectrum of radiated energy as:
\[\frac{\diff I(\omega)}{\diff \Omega} = \frac{e^2\omega^2}{16\pi^3}\left|\bd{\hat{r}} \times \int\displaylimits_{0}^{\infty} d t\, \bd{\beta}(t) \textrm{exp}[-i(\bd{k}\cdot \bd{r}_0(t) - \omega t)]\right|^2.\]

Now we make the approximation of infinite acceleration.  As can be seen, the integral was already set to zero over $(-\infty,0)$. The non-zero contribution is inside $(0,+\infty)$.  The electron is moving after $t>0$, with $\bd{\beta}(t) = \bd{\beta}_\textrm{f}$.  The trajectory function is $\bd{r}_0(t) = \bd{\beta}_\textrm{f} t$. Using $\bd{k} = \omega \bd{\hat{r}}$, 
\[\frac{\diff I(\omega)}{\diff \Omega} = \frac{e^2\omega^2}{16\pi^3}\left|\bd{\hat{r}} \times \bd{\beta}_\textrm{f}\right|^2\left|\int\displaylimits_{0}^{\infty} d t\, \textrm{exp}[-i\omega(\bd{\hat{r}}\cdot \bd{\beta}_\textrm{f}-1)t)]\right|^2.\label{omega2}\]
The prefactor $\omega^2$ in the frequency dependence will ultimately cancel out after integration. The integral diverges at late times (upper limit), so a convergence regulator $e^{-\epsilon t}$ is applied and $\epsilon \rightarrow 0$ after integration. Using $\bd{\hat{r}}\cdot \bd{\beta}_\textrm{f} = s \cos\theta$, where $\theta$ is the angle between $\bd{\beta}_\textrm{f}$ and the observation point, the integral is:
\[\int\displaylimits_{0}^{\infty} d t\, \textrm{exp}[i\omega(1-s \cos\theta)t-\epsilon t)] =  \frac{i}{i \epsilon+\omega(  1-s \cos \theta) },\]
where now $\epsilon \rightarrow 0$.  The square of the integral is:
\begin{equation}\left|\int\displaylimits_{0}^{\infty} d t\, \textrm{exp}[-i\omega(\bd{\hat{r}}\cdot \bd{\beta}_\textrm{f}-1)t)]\right|^2 =  \frac{1}{\omega^2(1- s \cos \theta )^2},\ee
demonstrating that frequency dependence cancels out exactly.  Using $\bd{\hat{r}}\times \bd{\beta}_\textrm{f} = s \sin\theta$, the result is Eq.~(\ref{sd_approx}):
\begin{equation}\frac{\diff I(\omega)}{\diff \Omega} = \frac{e^2}{16\pi^3} \left(\frac{s \sin\theta}{1-s \cos\theta}\right)^2.\ee
The angular distribution of energy radiated per unit frequency and the total energy radiated per unit frequency are independent of frequency. 


\subsection{Three-dimensional Planck}
Let us consider, for a final alternative derivation, how a three-dimensional Planck spectrum may describe photons emitted by an accelerated electron. The identity, Eq.~(\ref{1D_law}), and the derivative property of the Fourier transform, yield the motion that radiates a three-dimensional Planck spectrum:
\be
\left| \mathcal{F}_\omega \left[ \partial_t^n W(e^{ t}) \right] \right|^2 = \frac{\omega^{2n-3}}{e^{2\pi \omega} - 1}.\ 
\label{3D_law}\ee
{\color{blue}{Here we have used the Fourier-derivative rule $\mathcal{F}_\omega[\partial_t^n f(t)]=(i\omega)^n\mathcal{F}_\omega[f(t)]$, so that each additional time derivative contributes a factor of $\omega^2$ to the squared modulus.}} We have also used natural units and dimensionless variables for simplicity ($\kappa=1$).  For three-dimensional Planck, we need $n=3$, so that $\omega^3$ appears in the numerator.  Thus, it is good to investigate the acceleration and corresponding velocity forms:
\be \ddot{z}(t) = \partial_t^3 W(e^{ t}), \quad \dot{z}(t) = \partial_t^2 W(e^{ t}).\ee
To capture the particle spectrum, we restrict our attention to the non-relativistic regime, characterized by a maximum velocity parameter $V$.  The acceleration and velocity equations of motion are thus:
\be \ddot{z}(t) = \frac{27}{4}V \partial_t^3 W(e^{ t}), \quad \dot{z}(t) = \frac{27}{4}V\partial_t^2 W(e^{ t}).\ee
Larmor's power formula, $P = (2\alpha/3)\ddot{z}^2(t)$, using the fine structure constant $\alpha$, can be integrated over time to give the total emitted energy:
\be E = \frac{81}{320} \alpha V^2,\label{3D_energy}\ee
while the spectrum formula, $I = (4\alpha/3) |\ddot{z}(\omega)|^2 $, gives the three-dimensional Planck formula,
\be I(\omega) = \frac{4\alpha}{3} \left(\frac{27}{4}V\right)^2 \frac{\omega^3}{e^{2\pi \omega}-1},\ee
whose integration over frequency $\omega$ gives the same total emitted energy, Eq.~(\ref{3D_energy}). 

Notice that the trajectory equation of motion leads to a radiated particle count for the three-dimensional spectrum, which is finite:
\be N = \int_0^\infty\frac{I(\omega)}{\omega}\; \diff{\omega} = \frac{243}{16}\frac{\zeta(3)}{\pi^3}\alpha V^2,\ee
unlike the one-dimensional infrared soft particle count.

\section{Conclusions}
\label{sec:concl}

{\color{blue}{We have explicitly derived a Planckian spectral distribution for the radiation from a moving point charge.}}  Using the classical energy spectrum, a relationship between acceleration and temperature was obtained in classical electrodynamics, based on a specific non-uniformly accelerated equation of motion. The calculation involved deriving the spectrum of radiation emitted by a charged particle along this non-uniformly accelerating trajectory, in contrast to the uniform acceleration of the Davies-Unruh effect. The term `thermal radiation' refers to the electromagnetic radiation emitted by a hot object, {\color{blue}{independent of statistical equilibrium}}, following Planck's law for the frequency distribution of photons, dependent on the object's temperature.

In other words, the radiation emitted by the accelerating charged particle has a temperature defined by Planck's law. This phenomenon results from the particle's acceleration via the constant of motion $\kappa$, Eq.~(\ref{peel}).  
Notably, we arrived at Eq.~(\ref{temp}) by applying classical and semi-classical physics, demonstrating that acceleration can lead to {\color{blue}{Planckian}} radiation. The calculations offer a classical perspective and an elementary derivation of the relationship between acceleration and temperature.

    
\section{Acknowledgements} 

Funding comes partly from the FY2024-SGP-1-STMM Faculty Development Competitive Research Grant (FDCRGP) no.201223FD8824 and SSH20224004 at Nazarbayev University in Qazaqstan. Appreciation is given to the ROC (Taiwan) Ministry of Science and Technology (MOST), Grant no.112-2112-M-002-013, National Center for Theoretical Sciences (NCTS), and Leung Center for Cosmology and Particle Astrophysics (LeCosPA) of National Taiwan University.        

\bibliography{main} 

\end{document}